# Rugged HBT Class-C Power Amplifiers with Base-Emitter Clamping

Xi Luo, Subrata Halder, and James C.M. Hwang

Lehigh University, Bethlehem, PA 18015, USA.

*Abstract* — **The ruggedness of HBT Class-C power amplifiers was improved by adding an anti-parallel diode to the amplifier input to limit the negative swing of the base-emitter voltage. The improved amplifier could withstand 3:1 instead of 2:1 mismatch in CW operation, and 2.5:1 instead of 1.5:1 mismatch in pulse operation. In contrast to other approaches with emitter ballast, active feedback, or electrostatic discharge protection circuits, the present approach is simple to implement and has negligible impact on overall amplifier output power, gain or efficiency.**

## I. INTRODUCTION

Most power amplifiers must be sufficiently rugged to withstand output mismatch when antenna impedance changes with the environment, for example [1]. Ruggedness is especially critical to Class-C power amplifiers, for which electric breakdown is always a concern since Class-C power amplifiers were first made of vacuum tubes. To improve the ruggedness of HBT power amplifiers, several approaches have been proposed including adding an emitter ballast resistor [2], or electrostatic discharge protection circuit to the amplifier input [3], as well as using a collector voltage sensing circuit to adjust the input signal [4], the collector bias voltage [5], [6] or current [7]. However, these approaches tend to degrade the amplifier gain and output power, add circuit complexity and power consumption, and are slow to respond in the case of digital feedback. This paper proposes a different approach by adding an anti-parallel diode to the HBT base-emitter junction that is simple to implement and does not sacrifice the overall amplifier performance.

## II. EXPERIMENTAL

The proposed approach was demonstrated monolithically on-chip in *n-p-n* InGaP/GaAs HBT technology through a commercial foundry [8]. The HBTs typically exhibit a forward current cut-off frequency $f_T$ of 40 GHz, a maximum frequency of oscillation $f_{MAX}$ of 60 GHz, an open-emitter collector-base breakdown voltage $BV_{CBO}$ of 30 V, an open-base collector-emitter breakdown voltage $BV_{CEO}$ of 15 V, and an open-collector base-emitter breakdown voltage $BV_{BEO}$ of 7 V.

In HBT Class-C amplifiers, the base-emitter voltage $V_{BE}$ can be the most negative when the collector-emitter voltage $V_{CE}$ is the most positive to cause collector-base avalanche breakdown [9]. By limiting the negative swing of $V_{BE}$, breakdown can be prevented and the amplifier can be more rugged. Fig. 1 shows three types of Class-C amplifier demonstrated with an HBT of eight emitter fingers, each finger 2-μm wide and 20-μm long.

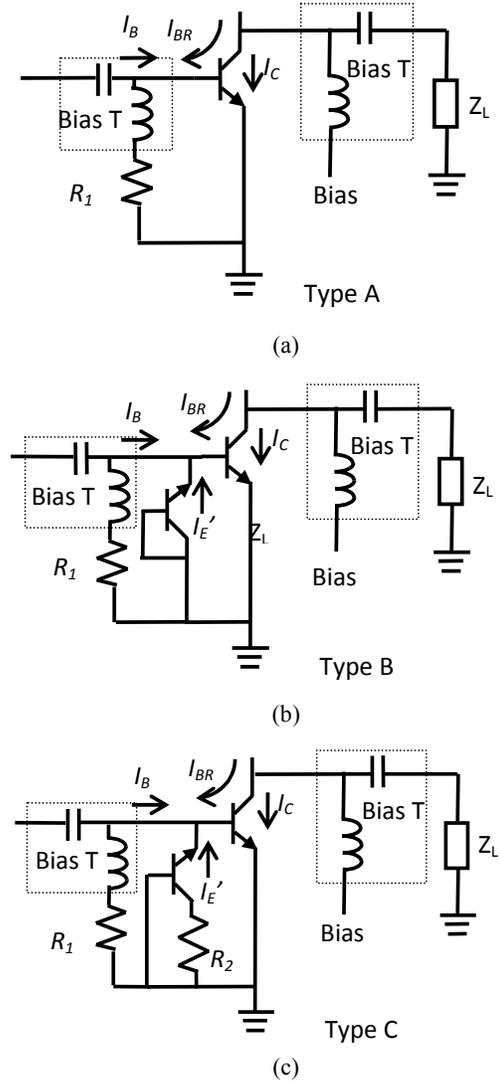

Fig. 1. Three types of HBT Class-C power amplifiers (a) without base-emitter clamping, (b) with base-emitter clamping, and (c) with base-emitter clamping enhanced by a series resistor $R_2$. The main HBT is self biased with resistor $R_2$ between the base and the emitter.

The HBT is self biased with a 50-Ω resistor $R_1$ between the base and emitter. In addition, in the Type-B and Type-C amplifiers, the based-emitter junction is shunted by a single-finger HBT of 2 μm × 20 μm. The single-finger HBT, with its base shorted to the collector, serves as an anti-parallel diode to limit the negative swing of $V_{BE}$ of the main HBT. Finally, in the Type-C amplifier, a series resistance $R_2$ of 100 Ω is added to the collector of the single-finger HBT to help drive it further into saturation under negative $V_{BE}$.

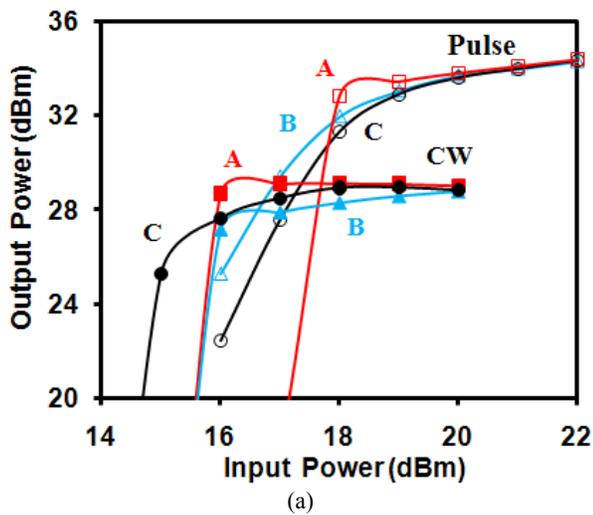

(a)

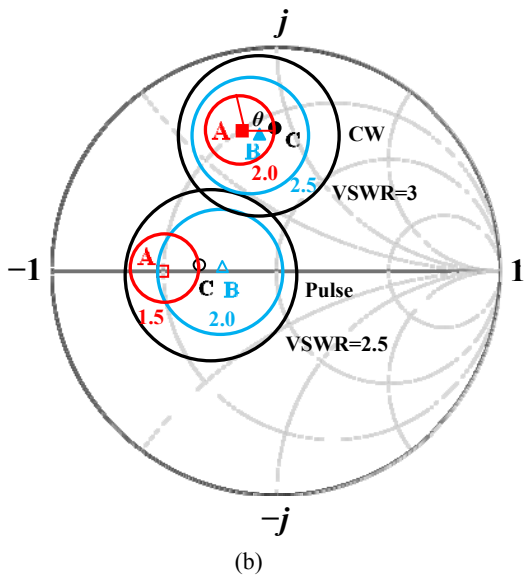

(b)

Fig. 2.  (a) Power-sweep characteristics and (b) optimum load impedances of (■, □) Type-A, (▲, △) Type-B, and (●, ○) Type-C amplifiers in (filled symbols) CW and (empty symbols) pulse operations. $V_{CE}$ = 12 V. θ represents mismatch angle. Maximum VSWR circles the amplifiers can withstand without burn out are plotted around the optimum loads.

In avalanche breakdown, holes generated in the collector can flow to the base ($I_{BR}$) and cause positive feedback and burn out. The base-emitter clamping diode in the Type-B or Type-C amplifiers can not only reduce the maximum collector-base voltage $V_{CB}$, but also draws current $I_E'$ into the base to compensate $I_{BR}$, thereby reducing the impact of avalanche breakdown.

The amplifiers were characterized at 6 GHz in both CW and pulse operations by using a Maury automated load-pull system. In pulse operation, both $V_{CE}$ and RF input power $P_{IN}$ were turned on for only 1 μs with a pulse repetition frequency of 2.78 kHz to minimize self heating. The RF output power was sampled in the middle of the pulse. The average collector current was sensed by an oscilloscope via a 5-Ω series resistor. In addition to power-sweep measurement in the frequency domain, time-domain current and voltage waveforms with three harmonics were measured at the base and collector by using an HP 71500A microwave transition analyzer [10].

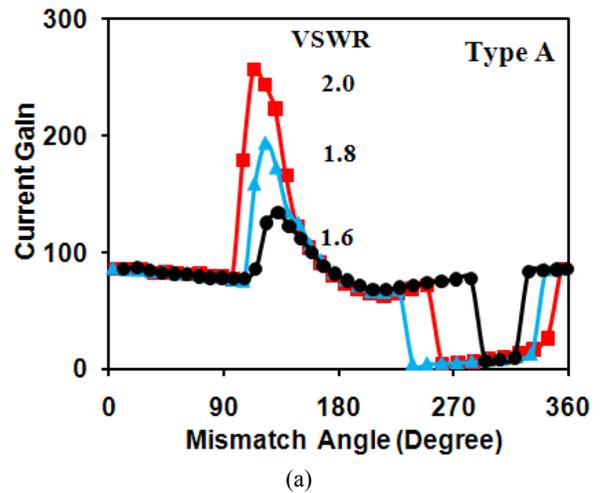

(a)

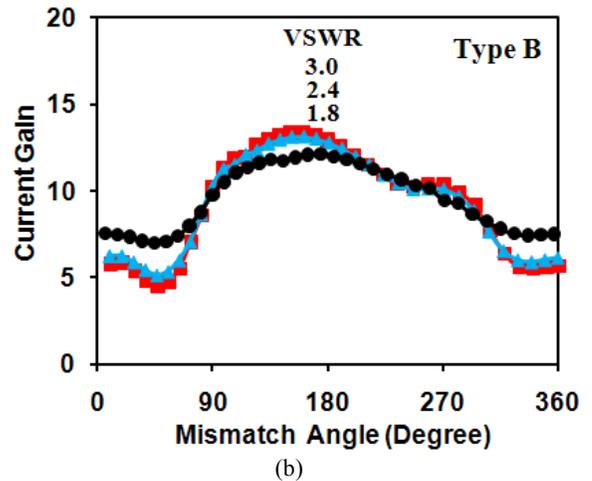

(b)

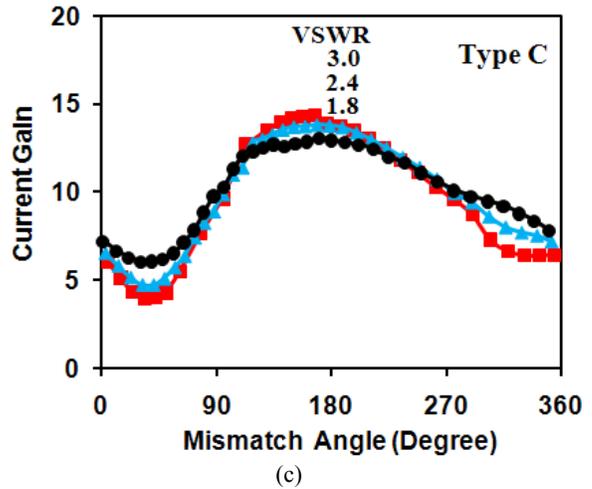

(c)

Fig. 3.  DC forward current gain for (a) Type-A, (b) Type-B, and (c) Type-C amplifiers in CW operation under different mismatch magnitude and angles. $V_{CE}$ = 12 V.

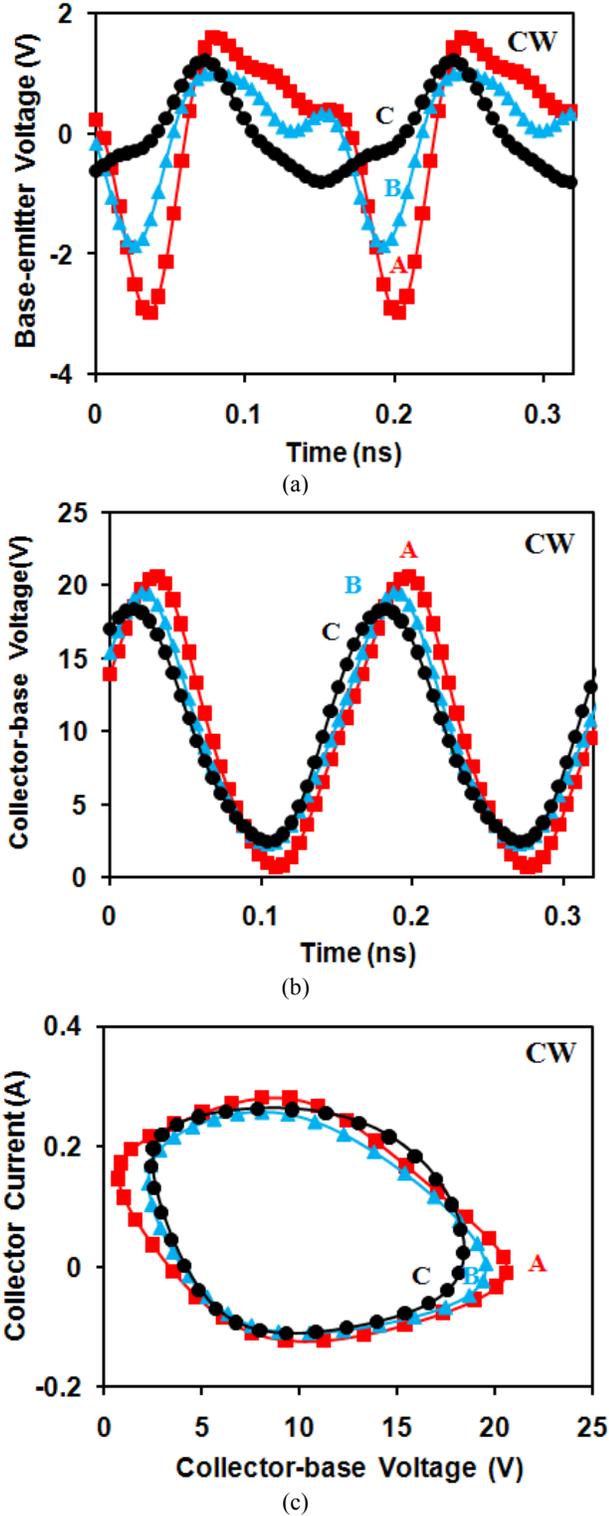

Fig. 4. (a) Base-emitter voltage waveform, (b) collector-base voltage waveform, and (c) collector dynamic load lines of (■) Type-A, (▲) Type-B, and (●) Type-C amplifiers in CW operation. $V_{CE}$ = 12 V. $P_{IN}$ = 14 dBm. $Z_L$ = 14 + j40 Ω.

## III. POWER-SWEEP CHARACTERISTICS

With $V_{CE}$ = 12 V and optimum source and load matches, the amplifiers delivered twice the output power in pulse operation than in CW operation. Fig. 2 shows that in CW operation, all three types of amplifier delivered 29-dBm output under optimum loads $Z_L$ of 22 + j44 Ω, 24 + j44 Ω, and 25 + j50 Ω for the Type-A, Type-B, and Type-C amplifiers, respectively. This shows that the modification for base-emitter clamping does not affect the amplifier output significantly. Similarly, in pulse operation all three types of amplifier delivered 34-dBm output, but the optimum loads became essentially real at 24 Ω, 35 Ω, and 37 Ω for the Type-A, Type-B, and Type-C amplifiers, respectively. The improvement in output power could be attributed to reduced Kirk effect and avalanche breakdown in pulse operation.

The amplifiers with base-emitter clamping were more rugged against load mismatch as determined by keeping the source impedance optimum while varying the load impedance for different voltage standing-wave ratios (VSWRs). It was found that in CW operation and under $V_{CE}$ = 13 V and $P_{IN}$ = 15 dBm, the maximum VSWRs the Type-A, Type-B, and Type-C amplifiers could withstand were 2.0, 2.5, and 3.0, respectively. Similarly, in pulse operation and under $V_{CE}$ = 10 V and $P_{IN}$ = 19 dBm, the maximum VSWRs were 1.5, 2.0 and 2.5 for the Type-A, Type-B, and Type-C amplifiers, respectively. The maximum VSWR circles were plotted around the optimum loads in Fig. 2(b).

Avalanche breakdown depends on not only the magnitude but also the angle θ of load mismatch. Fig. 3 shows the DC forward current gain in CW operation under different mismatch magnitudes and angles with $V_{CE}$ = 12 V and $P_{IN}$ = 14 dBm. It can be seen that for the Type-A amplifier, the current gain peaks between 110° and 160° of mismatch when the breakdown-generated electron current adds to the collector current while the breakdown-generated hole current subtracts from the base current. By contrast, the current gain of the Type-B or Type-C amplifier does not increase significantly with either the magnitude or angle of mismatch, indicating that avalanche breakdown is successfully suppressed.

## IV. CURRENT AND VOLTAGE WAVEFORMS

Because the current gain was difficult to measure directly in pulse operation, base and collector waveforms were measured to confirm the suppression of avalanche breakdown. Fig. 4 shows the base-emitter voltage waveform, collector-base voltage waveform, and dynamic load lines in CW operation with $V_{CE}$ = 12 V, $P_{IN}$ = 14 dBm and $Z_L$ = 14 + j40 Ω near where the current gain peaks. It can be seen that the most negative $V_{BE}$'s were approximately −3 V, −2 V and −1 V for the Type-A, Type-B and Type-C amplifiers, respectively. Meanwhile, the most positive $V_{CE}$'s were 21 V, 19 V and 18 V, for the Type-A, Type-B and Type-C amplifiers, respectively. The dynamic load lines confirm that the Type-A amplifier had the highest peak collector current and voltage.

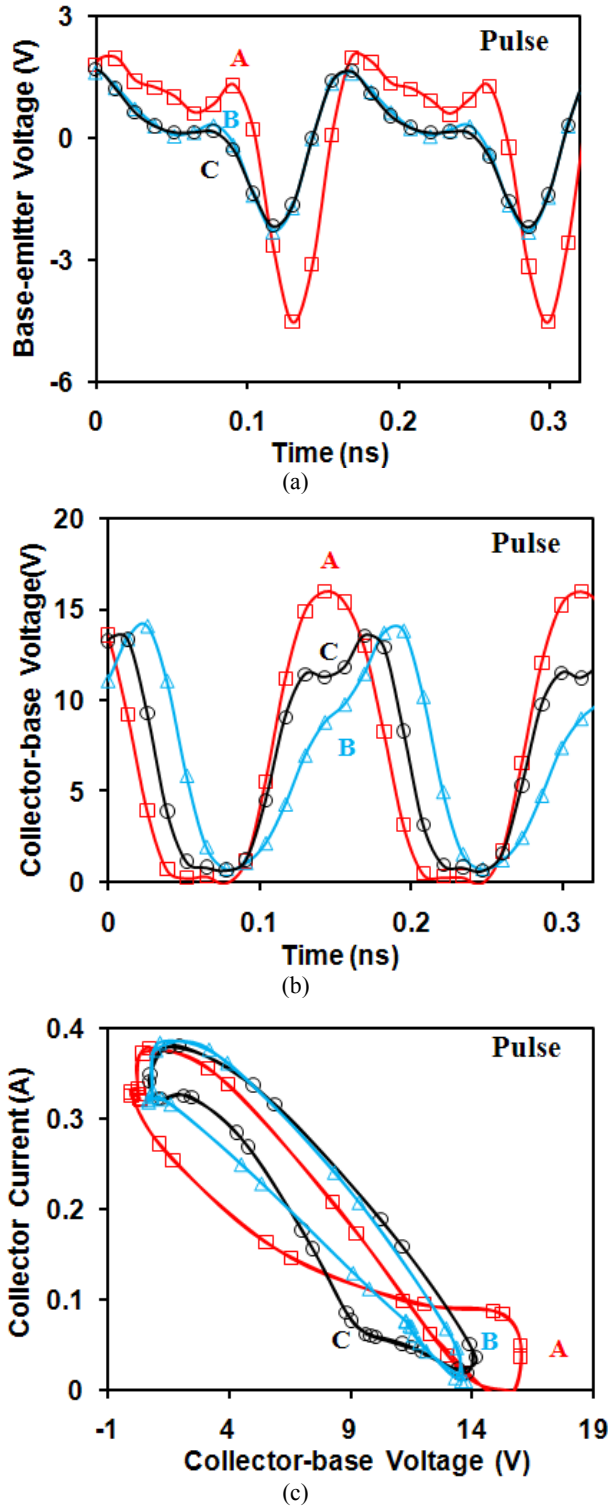

Fig. 5. (a) Base-emitter voltage waveform, (b) collector-base voltage waveform, and (c) collector dynamic load lines of (□) Type-A, (△) Type-B, and (○) Type-C amplifiers in pulse operation. $V_{CE}$ = 10 V. $P_{IN}$ = 19 dBm. $Z_L$ = 42 + j10 Ω.

The waveforms measured in pulse operation with $V_{CE}$ = 10 V, $P_{IN}$ = 19 dBm and $Z_L$ = 42 + j10 Ω exhibited the same trend as seen in Fig. 5.

## V. Conclusion

With base-emitter clamping on the input of HBT Class-C power amplifiers, avalanche breakdown as well as the impact of breakdown-induced hole current was suppressed. The improved amplifier could withstand higher load mismatch without sacrificing the overall amplifier output power, gain and efficiency.